\documentclass[12pt]{article}
\usepackage{amsmath,amssymb, amsthm,times,listings,hyperref,bm}
\usepackage[usenames,dvipsnames]{color}
\usepackage[pdftex]{graphicx}

\theoremstyle{definition}

\author{B. Cloutier\footnote{cloutbra@umich.edu, Dept. of Physics, University of Michigan}, B.K. Muite\footnote{muite@umich.edu, Dept. of Mathematics, University of Michigan}, P. Rigge\footnote{riggep@umich.edu, Dept. of Electrical Engineering and Computer Science, University of Michigan}}
\title{Performance of FORTRAN and C GPU Extensions for a Benchmark Suite of Fourier Pseudospectral Algorithms}

\begin{document}
\maketitle

\begin{abstract}
A comparison of PGI OpenACC, FORTRAN CUDA, and Nvidia CUDA pseudospectral methods on a single GPU and GCC FORTRAN on single and multiple CPU cores is reported.  The GPU implementations use CuFFT and the CPU implementations use FFTW. Porting pre-existing FORTRAN codes to utilize a GPUs is efficient and easy to implement with OpenACC and CUDA FORTRAN. Example programs are provided.
\end{abstract}

\section{Introduction}
Graphics processing units (GPUs) can have better performance for mathematical operations on large arrays when compared to traditional central processing units (CPUs). The fast Fourier transform (FFT) is one application for which GPUs have a significant performance advantage over CPUs. The performance advantage can be significant for simulations which fit within the memory constraints of a single GPU. 

GPU acceleration has largely been accomplished in variants of C with variants of FORTRAN being a recent addition. A comparison of performance of the lesser known OpenACC and CUDA FORTRAN on GPUs is of interest because a large number of legacy codes use FORTRAN, \cite{HenEtAl11} \cite{ZafGhoSebZho11} and because \cite[Chp.~6]{LevWag11} indicates that on CPUs, FORTRAN compilers typically generate more efficient scientific computing codes than C compilers. Previous works which have examined speedup offered by single GPUs in solving differential equations using Fourier transforms are \cite{BauKei11}, \cite{Che12} and \cite{ZhaWanYao09}.

In this paper, extensions of FORTRAN for GPUs when solving nonlinear PDEs with pseudospectral methods are compared.The user friendliness of these different GPU programming models is described. A benchmark suite of algorithms for three different nonlinear PDEs with Fourier pseudospectral methods is also provided.

\section{Programming Models}

At present it is unclear what programming model will be widely adopted for accelerators. Some current options include OpenCL, OpenACC, OpenHMPP, F2C-ACC, CUDA, and CUDA FORTRAN. This note will compare CUDA C, CUDA FORTRAN, and OpenACC GPU programs to serial and OpenMP\cite{OMPSpec12} FORTRAN CPU programs.

\subsection{FORTRAN and OpenACC}

OpenACC is a standard currently under development to allow for parallel programming of CPUs and GPUs\cite{OACCspec12}. It primarily uses directives to move data between a host CPU and a GPU and parallelize operations on the GPU.  FORTRAN, C, and C++ are supported and there is a hope that it will be merged with OpenMP in the future to allow for a unified interface to GPUs and other accelerators.

\subsection{CUDA C}

CUDA C is a set of extensions to C that allow special functions, called kernels, to be executed on supported NVIDIA GPUs\cite{NvCGuide12}. CUDA C provides a  lower level interface than many of the other application interfaces discussed here.

\subsection{CUDA FORTRAN}

CUDA FORTRAN, is a set of extensions to FORTRAN that allows kernels to execute on NVIDIA GPUs\cite{PgiFGuide12}. The main constructs are similar to CUDA C; however, CUDA FORTRAN has several directives that can automatically generate kernels for common cases. CUDA FORTRAN provides both a high level interface similar to OpenACC and a low level interface like CUDA C.

\section{An Overview of The Equations}

The three choosen equations, which are of mathematical and physical interest, can be solved entirely using Fourier pseudospectral methods and simple representative timestepping schemes. The resulting programs are short, can be understood in their entirety by a single person and only require an FFT routine.  

\subsection{The Cubic Nonlinear Schr\"{o}dinger Equation}\label{nlsoverview}

The focusing two dimensional cubic nonlinear Schr\"{o}dinger equation is
\begin{equation}
i\psi_t+\psi_{xx}+\psi_{yy}=\lvert \psi \rvert^2\psi, \label{eq:nls}
\end{equation}
where $\psi(x,y,t)$ is a complex valued function of time, $t$, and two spatial variables, $x$ and $y$. This equation arises in a variety of contexts including quantum mechanics,  in simplified models for lasers, and water waves. When $\psi$ has periodic boundary conditions, the cubic Schr\"{o}dinger equation has two conserved quantities,
\begin{equation}
\iint \lvert \psi \rvert^2\mathrm{d}x\mathrm{d}y \quad\textup{and}\quad
\iint \frac{\lvert \nabla \psi \rvert^2}{2}-\frac{\lvert \psi \rvert^4}{4}\mathrm{d}x\mathrm{d}y, \label{eq:nlsmassenergy}
\end{equation}
known as the mass and energy respectively; they can be used to assess the accuracy of a numerical solution. For more background on this equation, see \cite{SheTanWan11} and \cite{Yan10}.

\subsection{The Sine-Gordon Equation}\label{sgoverview}

The $2D$ sine-Gordon equation,
\begin{equation}
  u_{tt}-\Delta u = -\sin u, \label{eq:sg}
\end{equation}
arises in many different applications, including propagation of magnetic flux on Josephson junctions, sound propagation in a crystal lattice, and several others discussed in \cite{ScChMc73}. It has a conserved Hamiltonian
\begin{equation}
  H= \iint \frac{1}{2} \left( u_t^2 + |\nabla u|^2\right) + \left( 1 - \cos u\right)\mathrm{d}x\mathrm{d}y,
\end{equation}
which is useful in evaluating the accuracy of numerical solutions.

\subsection{The 2D Navier-Stokes Equation}\label{nsoverview}

The 2D incompressible Navier-Stokes equations in stream function ($\psi$)-vorticity ($\omega$) form are
\begin{equation}
 \omega_t+ \psi_y\omega_x -\psi_x\omega_y = \Delta \omega\quad\textup{and} \quad \Delta \psi = - \omega.
\end{equation}
These equations model fluid flow, and further background on them can be found in \cite{Tri88} among other references. The Taylor-Green vortex solution of these equations is
\begin{equation}
\omega(x,y,t)=4\pi\sin(2\pi x)\sin(2\pi y)\exp(-8\pi^2 t).  \label{eq:TaylorGreen}
\end{equation}

\section{Fourier Pseudospectral Methods}\label{numerics}

Fourier pseudospectral methods are a class of numerical methods to solve partial differential equations that utilize the Fourier transform. These methods became popular after the publication of \cite{GotOrs77}; other expositions are in \cite{Boy01}, \cite{CanEtAl06}, \cite{CheEtAl12},  \cite{HesGotGot07}, \cite{SheTanWan11}, \cite{Tre00} and \cite{Yan10}. These methods utilize the fact that differentiation of a function is a simple and fast multiplication by the wave number in Fourier space. The nonlinear terms in are computed in real space. This section gives a brief description of the second order in time algorithms used. The programs are available at \url{http://arxiv.org/abs/1206.3215}.

\subsection{A Numerical Method for the Nonlinear Schr\"{o}dinger Equation}\label{nlsnumerics}

The nonlinear Schr\"{o}dinger equation is approximated by splitting it into two equations which can be solved exactly, \cite{SheTanWan11} and \cite{Yan10}. The Fourier transform of $\psi$ will be denoted by $\hat{\psi}$. In Fourier space one first solves
\begin{equation}
i\psi_t+\psi_{xx}+\psi_{yy}=0 \label{eq:linSch}
\end{equation}
for half a time step, $0.5\delta t$. Letting $k_x$ and $k_y$ denote the wave numbers in the $x$ and $y$ directions, eq.\ \eqref{eq:linSch} is solved by $\hat{\psi}(t=0.5\delta t)=\exp[-i(k_x^2+k_y^2)\delta t]\hat{\psi}(t=0)$.  The solution of
\begin{equation}
i\psi_t=\lvert \psi \rvert^2\psi
\end{equation}
for a full time step is $\psi(t=\delta t)=\exp(i\lvert \psi(t=0) \rvert^2\delta t)\psi(t=0)$,  since $\lvert \psi \rvert^2$ is conserved. Finally  eq.\ \eqref{eq:linSch} is solved for another half time step to get the solution a full time step later. 

\subsection{A Numerical Method for the Sine-Gordon Equation}\label{sgnumerics}

Following \cite{DoSc10}, the second derivative in time is approximated by a central difference.  The resulting numerical method is
\begin{align}\label{eq:sgmethodunformatted}
&{}\frac{u^{n+1} -2u^n+u^{n-1}}{\delta t^2}+ \Delta\left( \frac{ u^{n+1}+2u^n+u^{n-1}}{4}\right) \notag
\\&{} = -\sin u^n
\end{align}

\subsection{A Numerical Method for the 2D Navier-Stokes Equation}\label{nsnumerics}

Time is discretized using the Crank-Nicolson method, where the nonlinear terms are solved for using fixed point iteration
{\small
\begin{align}
&{}\hspace{2em}\frac{\omega^{n+1,k+1}-\omega^n}{\delta t} - \frac{1}{2}\Delta\left(\omega^{n+1,k+1}+\omega^n\right) \notag
\\&{}+\frac{1}{2} \left( \psi_y^{n+1,k}\omega^{n+1,k}_x - \psi_x^{n+1,k}\omega^{n+1,k}_y
+ \psi_y^{n}\omega^{n}_x - \psi_x^{n} \omega^{n}_y  \right)=0. 
\end{align} }
The superscript $n$ denotes the time step and the superscript $k$ denotes the iterate. The fixed point iterations stop when
\begin{align}
\text{tolerance}>\iint(\omega^{n+1,k}-\omega^{n+1,k+1})^2\mathrm{d}x\mathrm{d}y.
\end{align}

\section{Results}\label{results}

\begin{table}[t]
\caption{A list of the compiler options used to build the codes. Codes in parentheses  use all flags in row. PGI 12.4 requires the flag -Mlarge\_arrays when more than 2Gb are allocated.}
\label{Table:info}
\centering
\begin{tabular}{|c|c|c|c|} \hline
                  & Compiler & Flags & Libraries   \\ \hline
&   & -O3      -Mcuda     & CUFFT   \\
GPU: Cuf and (OpenACC)   &  PGI 12.4   & -Minfo (-acc)  &  4.1.28  \\
&  & (-ta=nvidia) & \\
 \hline
& CUDA 4.0 & -O3  & CUFFT\\
GPU: C                &   V0.2.1221    & --arch=sm\_20&  4.1.28   \\
\hline
 & GFORTRAN  & -O3 & FFTW \\
CPU: 1 core and (16 core)       &     GCC: 4.6.2      &    (-fopenmp)    & 3.3   \\

\hline
\end{tabular}
\end{table}

\begin{table}[t]
\caption{Computation times in seconds for 20 time steps of $10^{-5}$ for a Fourier split step method for the cubic nonlinear Schr\"{o}dinger equation on $[-5\pi,5\pi]^2$.}
\label{Table:NLS}
\centering
\begin{tabular}{|c|c|c|c|c|c|} \hline
Grid & GPU & GPU & GPU & CPU  & CPU \\
 Size                 & (Cuf) &  (C) & (OpenACC) &(16 cores) &  (1 core)  \\
\hline
$64^2$ & 0.00292 & 0.00618 &0.00272&  0.0180 & 0.0180 \\
$128^2$ & 0.00366 & 0.007189 &0.00459& 0.0130 & 0.086\\
$256^2$ & 0.00802 & 0.0116 &0.0130& 0.148 & 0.442\\
$512^2$ & 0.0234 & 0.0315 &0.0369& 0.562 & 1.94\\
$1024^2$ & 0.0851 & 0.105 &0.132& 2.27 & 12.7\\
$2048^2$ & 0.334 & 0.415 &0.527& 9.67 & 57.2 \\
$4096^2$ & 1.49 & 2.02 &2.30& 37.7 & 329 \\
$8192^2$ & 6.30 & N/A &N/A& 292.4 &1454 \\
\hline
\end{tabular}
\end{table}%

\begin{table}[t]
\caption{Computation times in seconds for 500 time steps of $10^{-3}$ the sine-Gordon equation on $[-5\pi,5\pi]^2$. N/A implies that the code could not be run due to memory constraints.}
\label{Table:SG}
\centering
\begin{tabular}{|c|c|c|c|c|c|} \hline
Grid & GPU & GPU & GPU & CPU  & CPU  \\
Size & (Cuf) & (C) &  (OpenACC) & (16 cores)	 & (1 core) \\
\hline
$64^2$   & 0.028  & 0.025  & 0.028  & 0.050  & 0.098  \\
$128^2$  & 0.040  & 0.031  & 0.041  & 0.107  & 0.401  \\
$256^2$  & 0.099  & 0.065  & 0.095  & 1.960  & 1.899  \\
$512^2$  & 0.343  & 0.260  & 0.344  & 2.925  & 9.301  \\
$1024^2$ & 1.148  & 0.976  & 1.218  & 19.07  & 38.34  \\
$2048^2$ & 4.485  & 4.007  & 4.869  & 67.84  & 165.9  \\
$4096^2$ & 18.09  & 16.53  & 19.95  & 481.2  & 785.5  \\
$8192^2$ & 85.45  & N/A    & 93.51  & 934.2  & 4556   \\
\hline
\end{tabular}
\end{table}%

\begin{table}[t]
\caption{Computation times in seconds for 20 time steps of $1.25\times10^{-3}$ for the incompressible Navier-Stokes equation on  $[0,1]^2$.  Each simulation generated the same max error of $1.65\times10^{-5}$ when compared to the exact Taylor-Green solution. The tolerance for fixed point iterations was set to $10^{-10}$. }
\label{Table:NS}
\centering
\begin{tabular}{|c|c|c|c|c|c|} \hline
Grid & GPU & GPU & GPU & CPU & CPU  \\
Size & (Cuf) & (C) & (OpenACC) &  (16 Cores) & (1 core) \\
\hline
$64^2$ &  0.0151 & 0.0164 & 0.0171& .0018 &0.014  \\
$128^2$ & 0.0201 & 0.0266  & 0.0204	& 0.024   &0.065 \\
$256^2$ & 0.044 & 0.0435 & 0.0418 & 0.417 & 0.37  \\
$512^2$ & 0.119 & 0.141 & 0.1205 & 0.939 & 1.734 \\
$1024^2$ & 0.357 & 0.520 & 0.398 & 4.89 & 7.65  \\
$2048^2$ & 1.386 & 1.88 & 1.59 & 17.04 &33.56  \\
$4096^2$ & 5.814  &  7.48 & 6.79&112.83 &172.67   \\
\hline
\end{tabular}
\end{table}

All three equations are solved using different numerical methods so actual computation time comparisons between methods are of less interest than common scaling trends. For this comparison all of the codes were compiled with the compilers and flags shown in Table \ref{Table:info}. \\

\begin{figure}
\caption{Comparison of OpenACC benchmark time to CuF, CUDA C, and CPU times for
a grid size of $4096^2$ (bigger is better).}
\label{Fig:Chart}
\centering
\includegraphics[width=3.5in]{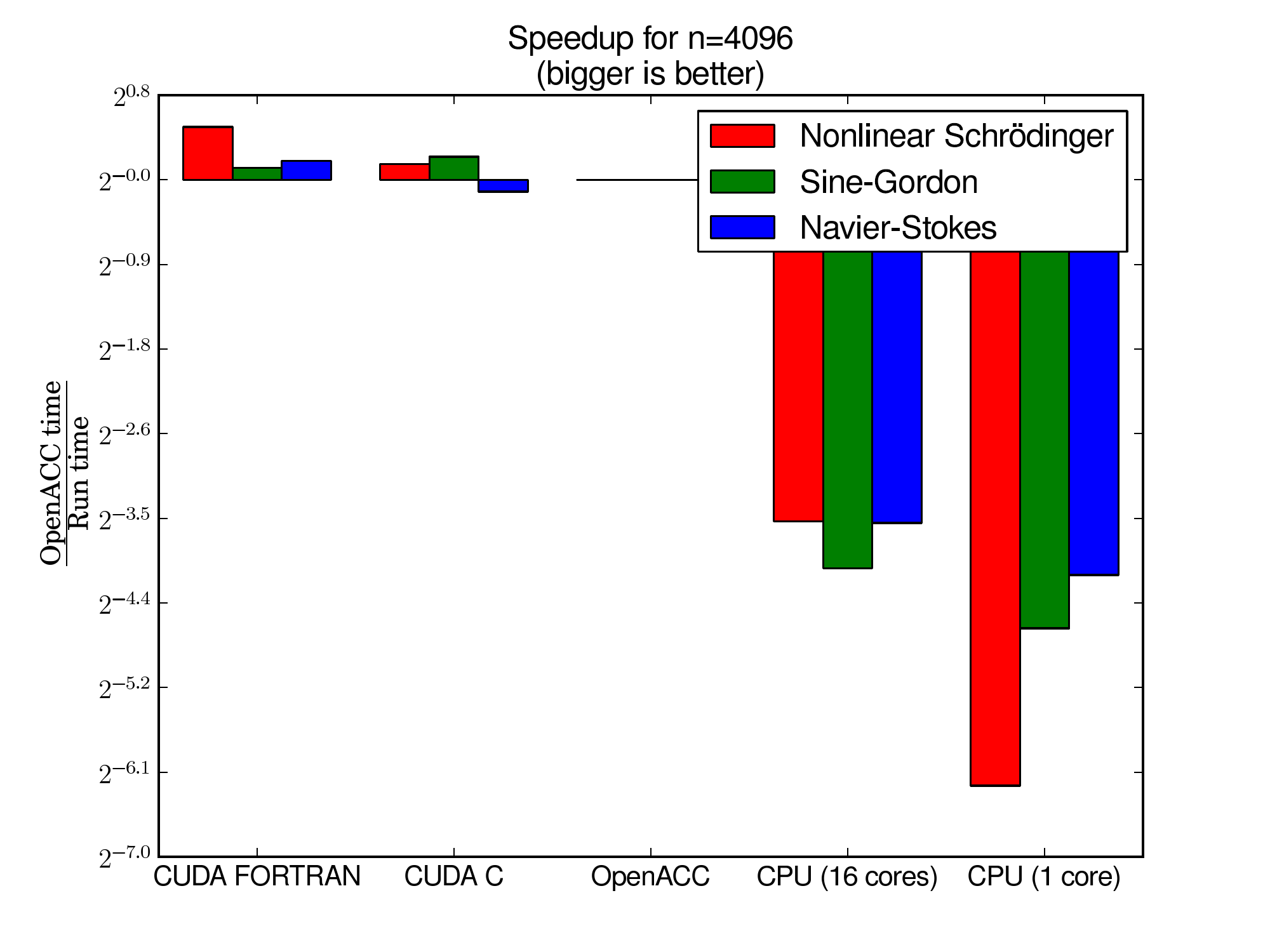}
\end{figure}

The CPU simulations were run on  AMD Opteron Magny-Cours 6136 2.4 GHz dual-socket eight-core processors with 48GB of memory and the GPU simulations on Nvidia Fermi M2070 with 6GB of memory per GPU.  Double precision floating point arithmetic was used. Computation times are only measured for advancing the numerical solution forward in time, they do not include setup time for creating FFT plans, allocations, calculating exact solutions, etc. because in production simulations where many time steps are taken, these will be negligible. Performance differences between FORTRAN compilers on CPUs do not change the conclusions.

\subsection{The Cubic Nonlinear Schr\"{o}dinger Equation}\label{nlsresults}

The programs used 4 double complex arrays, 2 arrays for the actual computation and 2 further arrays, to calculate the mass and energy. In all tests, the mass was conserved up to machine precision and the energy was constant to within 6 significant figures.

Table \ref{Table:NLS} shows that the GPU gives a speed up of up to a factor of 40 compared to a multicore OpenMP CPU implementation. An initial naive CUDA FORTRAN implementation for which kernel parallelization options are left to the compiler was typically a factor of 2 slower than the CUDA C implementation for which block sizes were specified. GPUs had speed ups of order 100 times compared to a single CPU core. Figure \ref{Fig:Chart} shows that for this code, CUDA Fortran had the best performance.

\subsection{The Sine-Gordon Equation}\label{sgresults}

The programs used real-to-complex Fourier transforms. They required 2 double precision arrays of size $N_x\times N_y$ and 3 double complex arrays of size $N_x\times(N_y/2+1)$. In all tests, the final Hamiltonian was within 6 significant figures.

Table \ref{Table:SG} shows that CUDA C performs the best, followed closely by CUDA FORTRAN and OpenACC. The differences between the GPU implementations were relatively small compared to the difference between GPU and CPU implementations. The GPU implementations performed on the order of 50 times better than a single core CPU.  Figure \ref{Fig:Chart} shows that for this code, CUDA C had the best performance.

\subsection{The 2D Navier-Stokes Equation}\label{nsresults}

The programs use the same real-to-complex Fourier transforms as the sine-Gordon equation, where the real-valued input array has a size of $N_x \times N_y$ and the complex-valued output array is of size $N_x \times (N_y/2+1)$. 10 arrays are used for the GPU codes and 11 arrays are used for CPU codes, where the extra array is required because FFTW does not preserve its input. Five complex-real transforms and 1 real-complex transform are used in the timestepping loop. 

Table \ref{Table:NS} shows that the GPU gives a speed up of up to a factor of 30 compared to a single CPU core. Figure \ref{Fig:Chart} shows that for this code, OpenACC does better than CUDA C, and CUDA FORTRAN has the best performance.

\section{User Comparison of the different GPU programming environments}
Porting existing FORTRAN codes to CUDA FORTRAN was simple and intuitive; the combination of FORTRAN and Open ACC was a little less intuitive, but still relatively straightforward. The differences between CPU and GPU versions with CUDA FORTRAN or OpenACC are small, so it is feasible to maintain CPU and GPU versions of the same code. In contrast, porting a FORTRAN code to CUDA C is error prone, requires significant reprogramming and a good understanding of the GPU architecture.  

Finally, \cite{HenEtAl11} observed that the F2C-ACC directive based FORTRAN to CUDA compiler results in a runtime code with better performance than regular CUDA FORTRAN.  The performance of CUDA FORTRAN codes may experience very different speedups with minor changes to the parallelization or compiler options. Choosing these options optimally requires some knowledge of the underlying architecture -- autotuning compilers would be useful for doing this.

\section*{Acknowledgment}

This work used the computers Forge and Kollman within the Extreme Science and Engineering Discovery Environment, which is supported by National Science Foundation grant number OCI-1053575.  B. Cloutier  and P. Rigge were supported by the University of Michigan Undergraduate Research Opportunity Program and Blue Waters Undergraduate Petascale Education Program respectively. The authors acknowledge support from a faculty grant for innovations in teaching with technology from the University of Michigan and a Blue Waters Undergraduate Petascale Module Development grant from the Shodor Foundation. They thank Bennet Fauber, Robert Krasny, Carl Ponder, Sarah Tariq, Divakar Viswanath, Jared Whitehead and Brian Wylie for help and suggestions, and the reviewers for a careful reading.

\end{document}